\documentstyle[floats,twocolumn,epsfig,aps]{revtex}

\def\>{\rangle}
\def\<{\langle}
\def\I{\mid}

\def\ket#1{\I #1 \>}
\def\braket#1#2{\< #1 \I #2\>}
\def\P#1{{\hat P}_{#1}}

\def\be{\begin{equation}}
\def\ee{\end{equation}}
\def\ba{\begin{eqnarray}}
\def\ea{\end{eqnarray}}

\begin{document}

\begin{titlepage}
\date{\today}
\title{NONLOCAL EFFECTS OF PARTIAL MEASUREMENTS \\AND QUANTUM ERASURE}

\author{Avshalom C. Elitzur\footnote{email: cfeli@weizmann.weizmann.ac.il}}
\address{Unit of Interdisciplinary Studies, Bar-Ilan University, 52900 Ramat-Gan, Israel.}
\address{and}
\address{The Bhaktivedanta Institute, Juhu, Juhu Road, Mumbai 400049, India.}

\author{Shahar Dolev\footnote{email: shahar\_dolev@email.com}}
\address{The Kohn Institute for the History and  Philosophy of Sciences, \\ Tel-Aviv University, 69978 Tel-Aviv, Israel.}
\maketitle

\begin{abstract}

Partial measurement turns the initial superposition not into a definite outcome but into a greater probability for it. The probability can approach 100\%, yet the measurement can undergo complete quantum erasure. In the EPR setting, we prove that i) every partial measurement nonlocally creates the same partial change in the distant particle; and ii) every erasure inflicts the same erasure on the distant particle's state. This enables an EPR experiment where the nonlocal effect does not vanish after a single measurement but keeps ``traveling" back and forth between particles. We study an experiment in which two distant particles are subjected to interferometry with a partial ``which path" measurement. Such a measurement causes a variable amount of correlation between the particles. A new inequality is formulated for same-angle polarizations, extending Bell's inequality for different angles. The resulting nonlocality proof is highly visualizable, as it rests entirely on the interference effect. Partial measurement also gives rise to a new form of entanglement, where the particles manifest correlations of multiple polarization directions. Another novelty in that the measurement to be erased is fully observable, in contrast to prevailing erasure techniques where it can never be observed. Some profound conceptual implications of our experiment are briefly pointed out.
\vspace{1cm}


\end{abstract}


\end{titlepage}


\section*{Introduction}               

Bell's theorem \cite{Bel} has made it possible, for the first time, to experimentally demonstrate quantum nonlocality. Later, the GHZ experiment \cite{Gre} and Hardy's \cite{Har} proof without inequalities extended the proof to new domains. All these proofs, however, involve complete measurements. This is insufficient since, at the quantum level, measurement can be a continuous process, the intermediate stages of which have seldom been studied. 

Is it possible to prove that nonlocal effects are produced even by small stages of the measurement process? Moreover, is it possible to show that nonlocal effects are formed not only by measurement but also by the time-reversed process, namely, quantum erasure? Affirmative answers would render nonlocality much more intriguing because, in the ordinary EPR experiments, a single measurement of a particle disentangles it, and no further measurements can reveal nonlocal effects. Once, however, the above two questions are answered in the affirmative, nonlocality will turn out to connect not only discrete events but continuous {\it processes} as well. The nonlocal influence will then appear to ``bounce" back and forth, many times, between the distant particles during the measurements. Several other intriguing features of QM, such as reversibility and information capacity, would also become manifest. 

The organization of this article is as follows. Section I introduces interaction-free measurement. Section II shows how, when the wave function is appropriately split, interaction-free measurement becomes partial. Section III shows that such a measurement obeys the uncertainty relations in that it partially disrupts a non-commuting variable. Section IV shows that partial measurement(PM) can sometimes be completely reversed. Section V presents a hybrid, EPR-PM experiment in order to show that partial measurement and its erasure exhibit nonlocal effects. Section VI shows that quantum theory's predictions for this experiment quantitatively differ from those of a local theory. In Section VII a Bell-like inequality is shown to break by the EPR-PM experiment. Section VIII studies the case of a multiple partial measurement on one particle and its unique nonlocal consequences. Section IX discusses the novel form of quantum erasure made possible by our experiment. Section X points out some bearings of these findings on quantum theory.

\section { Interaction-Free Measurement and the Uncertainty Relations}

Single-particle interferometry provides some of the most intriguing illustrations for quantum-mechanical principles, and in recent years it has become technically feasible. Consider a photon entering a calcite crystal positioned to divert the incident photon according to its polarization along the $x$ axis (Fig.\ 1). If the photon's polarization is $90^{\circ}$ ($P_x=+1$, $\ket{\uparrow}$), it will be diverted to the lower path, whereas if it is $0^{\circ}$ ($P_x=-1$, $\ket{\rightarrow}$) it will be diverted to the upper path. The calcite thus acts as a polarizing beam splitter (pBS), similar to the Stern-Gerlach magnet for spin-$^1\!/_2$ particles. As long as no measurement has been made to find out which path the photon took -- thereby leaving the photon's polarization undetermined too -- the photon will remain in a superposition of both paths and both polarizations:
\be
\ket{\!\Psi}=c_1\!\ket{\uparrow}+c_2\!\ket{\rightarrow}.
\ee
So far, the splitting of the wave function is reversible. A second calcite, aligned at the $x$ plane as the first but facing an opposite direction (denoted by $\bar{x}$), re-unites the resulting $\ket{\uparrow}$ and $\ket{\rightarrow}$ beams, so that the photon re-emerges in one single beam, in the same state as it has entered the first calcite (Fig.\ 1).

\begin{figure}[t]
\centering
\includegraphics{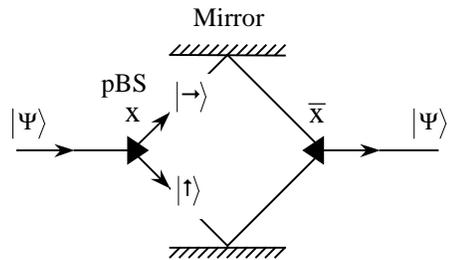}
\medskip
\caption{ It is possible to restore photon's exact state as long as no measurement is performed.}
\end{figure}

The reversibility of splitting the photon along the $x$ direction is demonstrated by the measurement of another, noncommuting variable. Before splitting, let the photon impinge on a calcite positioned in $45^{\circ}$ to the $x$ direction, measuring its $y$ polarization. Suppose that the $y$ polarization has been found to be $+45^{\circ}$ ($P_y=+1$, $\ket{\nearrow}$). Then, let the photon split according to its $x$ polarization and re-unite again (Fig.\ 2). If no measurement has been made between the splitting and re-uniting, we are left ignorant about the photon's $x$ polarization. Consequently, the $y$ polarization will remain intact: a final $y$ measurement will {\it always} yield a polarization of $+45^{\circ}$, just like the initial $y$ measurement. 

\begin{figure}[t]
\centering
\includegraphics{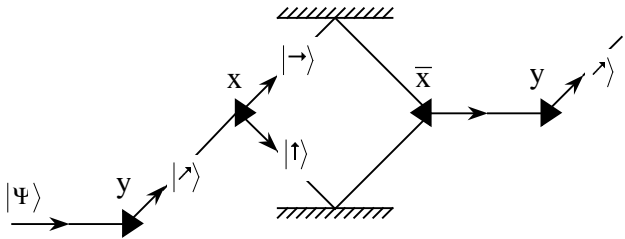}
\medskip
\caption{ As long as no measurement is taken to tell the photon's polarization in the $x$ direction, its polarization along the $y$ direction remains intact.}
\end{figure}

Suppose, however, that two detectors are placed on the two routes of the split wave function prior to its reunification (to enable the later reunification of the rays, let the measurements be of the non-demolition type, such that the detectors do not absorb the photon in case of detection). Since the $\ket{\nearrow}$ state is an even mixture of $\ket{\rightarrow}$ and $\ket{\uparrow}$,
\be
\ket{\nearrow}=\frac{1}{\sqrt{2}}(\ket{\uparrow}+\ket{\rightarrow}),
\ee
in 50\% of the cases the upper detector will click, and in the other 50\% the lower one will. This measurement has an irreversible consequence: The knowledge we have gained about the photon's polarization will takes the cost of blurring the other, non-commuting variable, obeying an uncertainty relation of the form:
\be
\Delta P_x \cdot \Delta P_y \ge 2i[P_x,P_y].
\label{eq:uncer}
\ee

Fig.\ 3 demonstrates this principle. Once $P_x$ is ascertained, $P_y$ is disrupted; a second $y$ measurement will yield either $+45^{\circ}$ ($\ket{\nearrow}$) or $-45^{\circ}$ ($\ket{\searrow}$), randomly. This is similar to the ordinary interference effect in a Mach-Zender Interferometer (MZI), where obtaining which-path information disrupts the photon's initial momentum. 

\begin{figure}[t]
\centering
\includegraphics{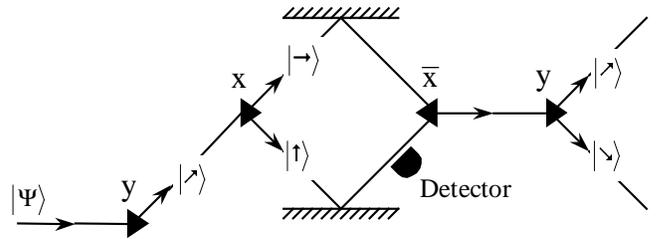}
\medskip
\caption{ A measurement that discloses the photon's $x$ polarization destroys the $y$ polarization even when the measurement has been carried out by a single detector that did not click.}
\end{figure}

Here a peculiar possibility emerges: What happens if, in order to measure $P_x$, we place only {\it one} detector on one of the two routes?  If the detector clicks, the answer is clear: This is an ordinary measurement, hence we should observe the above disruption of $P_y$. The situation becomes more intriguing in the remaining 50\% of the cases, when the single detector clicks not. Although no observable interaction has taken place, the very silence of the detector indicates that the photon has traversed the other path, thereby disclosing the photon's polarization with certainty. Hence, $P_y$ should be disrupted in this case too, just because the silent detector {\it could} have clicked!

This ``interaction free measurement" \cite{Eli} has been the subject of intense experimental and theoretical study in the last few years \cite{Har2,Kwi,Pen}(see \cite{Lev} for a comprehensive review). In the present context, it has two unique features that warrant attention. First, with a simple modification, interaction-free measurement can be partial, leaving the wave function in a certain degree of superposition even after the measurement. Second, it can be completely reversed.

\section { Measurement can be Partial }
\label{se:partial}

Nearly always, measurement is regarded as a single event, whereby the superposition of all possible states gives its place to one state. In reality, however, there can be many intermediate stages in the measurement process, stages that only change the initial probabilities without yet giving a definite result. \cite{Woo,Dor,HB1,HB2}

Following is a simple apparatus for partial measurements. Let a photon be split by a calcite in the above manner. Next, with the aid of many partly-silvered mirrors, let the $\ket{\uparrow}$ beam split further into 100 beams. The transmission coefficients of the mirrors are graded such that all the 100 beams have an equal intensity of 1\% of the original beam (Fig.\ 4). The first mirror transmits $\frac{99}{100}$ of the incident beam, the second $\frac{98}{99}$ and so on, with the $99^{th}$ mirror transmitting $^1\!/_2$ of the beam, and the last one being a solid mirror. Similarly, let the $\ket{\rightarrow}$ beam split into another 100 by the same technique. The photon is now in a superposition of 200 beams, such that, if its polarization along the $x$ direction is $90^{\circ}$ it might be detected in one out of the lower 100 beams, whereas if its polarization is $0^{\circ}$ it might be detected in one out of the 100 upper beams.

Finally, let a complementary series of mirrors re-unite all 200 beams into two $\ket{\uparrow}$ and $\ket{\rightarrow}$ beams, and let a reversed $x$ calcite re-unite the resulting beams. The apparatus keeps the paths of all the beams in the same length, so as to keep the beams in phase.\footnote{This is, essentially, a modification of the Mach-Zender interferometer. Aligning these mirrors to about one wavelength error is a difficult task, but completely possible in current technology.} This way, the entire splitting process is reversed and the photon re-emerges as one single beam. 

Here too, we can demonstrate the superposition of the photon's $x$ polarization by performing two $y$ polarization measurements, one before and one after the splitting and re-unification processes.

As long as no measurement is performed, the superposition will remain intact and all the photon will always emerge in the same state as it entered the apparatus, e.g., $\ket{\nearrow}$ as in Fig.\ 4.

\begin{figure}[htbp]
\onecolumn\twocolumn[
\centering
\includegraphics{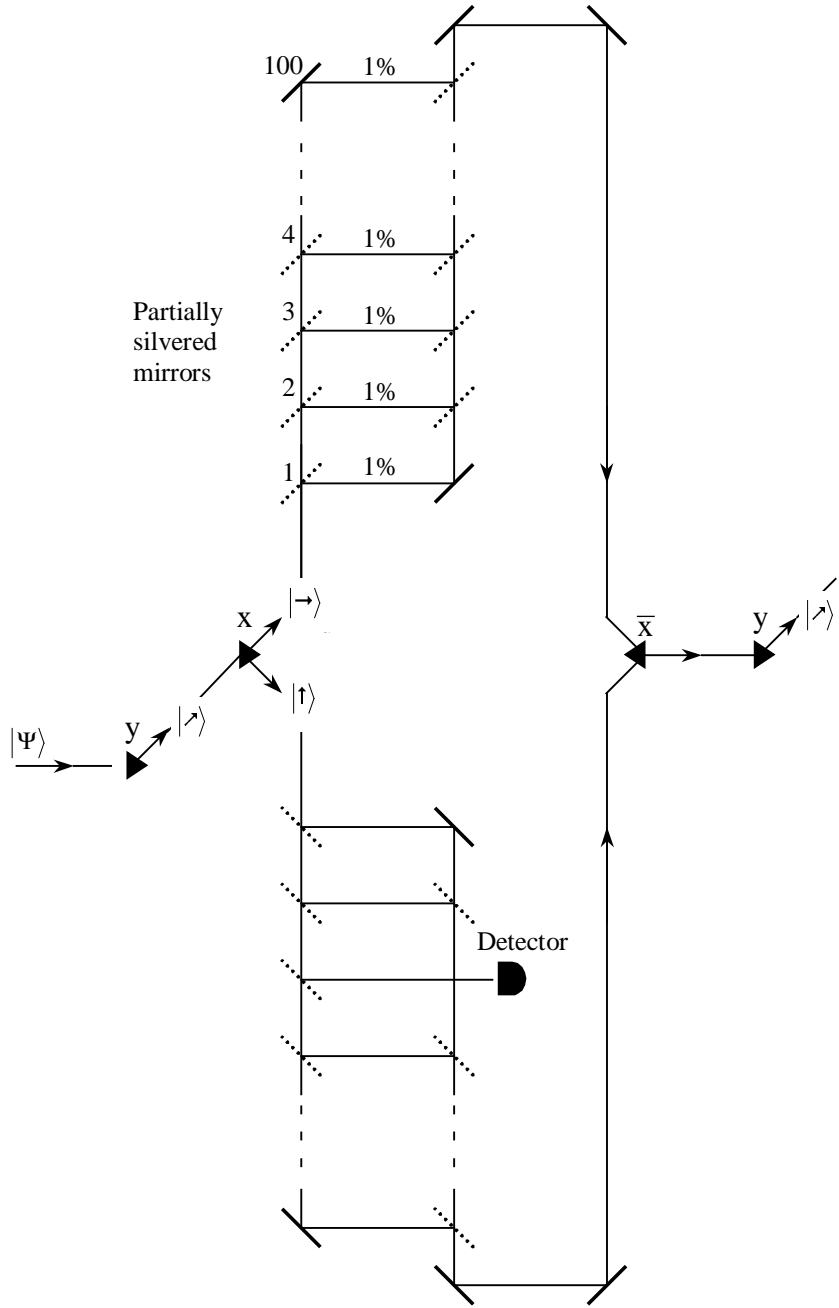}
\onecolumn
\caption{ A series of partly-silvered mirrors splits each half of the wave function into 100 equal beams. Here too, as long as no measurement has been taken on any of the 200 beams, the $y$ polarization remains intact.}
\twocolumn
]
\end{figure}

This setup enables performing partial measurements. We can now remove one of the mirrors and place a detector, as demonstrated on the third beam of the $\ket{\uparrow}$ branch of Fig.\ 4. If the detector clicks, we know for certain that the photon's polarization is $\ket{\uparrow}$ and the measurement is complete. In 99.5\% of the cases, however, the detector will not click. This constitutes an interaction-free measurement, which should alter the wave function. Yet, since the portion of the wave function thus measured is so tiny, the superposition will change only slightly,
\be
\ket{\nearrow}=\frac{1}{\sqrt{2}}(\ket{\uparrow}+\ket{\rightarrow}) \; \longrightarrow \; \sqrt{\frac{99}{199}}\ket{\uparrow}+\sqrt{\frac{100}{199}}\ket{\rightarrow},
\ee
and will continue slightly changing for each further measurement on the remaining beams that yields no click. In other words, each interaction-free measurement slightly reduces the probability that the photon's polarization is $\ket{\uparrow}$, thereby increasing its probability to have a $\ket{\rightarrow}$ polarization. For $n$ detectors on the $90^{\circ}$ branch, the effect on the wave-function will be
\ba
&&\frac{1}{\sqrt{2}}(\ket{\uparrow}+\ket{\rightarrow}) \nonumber\\
&&\quad\longrightarrow\sqrt{\frac{100-n}{200-n}}\ket{\uparrow}+\sqrt{\frac{100}{200-n}}\ket{\rightarrow}\nonumber\\
&&\quad=\sqrt{\frac{\alpha}{1+\alpha}}\ket{\uparrow}+\sqrt{\frac{1}{1+\alpha }}\ket{\rightarrow},
\label{eq:measn}
\ea
where $\alpha$ is the intensity of the (unmeasured) $\ket{\uparrow}$ beam:
\be
\alpha=\frac{100-n}{100}.
\label{eq:alpha}
\ee
Let us denote the operator of partial polarization measurement in the $x$ up direction with intensity $\alpha$ as:
\be
\hbox{Partial Polarization Measurement}\equiv \P{\uparrow \alpha}. \quad \quad
\ee
(Note again that $\alpha$ is the unmeasured intensity).\\
This operator obeys the following multiplication law:
\be
\P{\uparrow\beta}\cdot \P{\uparrow\alpha}=\P{\uparrow\alpha\cdot\beta}.
\ee

A familiar question, central to quantum theory, now poses itself: {\it Does partial measurement change merely our knowledge about the photon or is this a real physical change going on with the photon's state?} Interaction-free measurement provides a straightforward way to show that the latter is true.

Recall that, prior to the photon's splitting along the $x$ direction, its polarization has been measured along the $y$ direction and was found to be $\ket{\nearrow}$.  All one has to do now is to reunite the $\ket{\uparrow}$ and the $\ket{\rightarrow}$ beams, and then measure again the $y$ polarization. If no attempt has been made to determine which of the 200 paths the photon took in the $x$ direction, the photon will emerge from the interferometer with its $x$ polarization unmeasured, hence its $y$ polarization will remain $\ket{\nearrow}$ with 100\% certainty. If, however, a measurement has been made to find out the photon's $x$ polarization, this measurement will have a proportionate effect. If one of the detectors click, then the measurement is complete and the photon's $y$ polarization will be totally disrupted, as in Fig.\ 3. But if the measurement is interaction-free -- i.e., a few beams measured with no click -- the $y$ polarization will be only partly disrupted (see also Ref.\ \cite{Woo}). Instead of a pure $\ket{\nearrow}$ state, we will have:
\ba
\ket{\!\Psi_\alpha}&\equiv&\P{\uparrow\alpha}\ket{\!\Psi}\nonumber\\
&=&\sqrt{\frac{\alpha}{1+\alpha}}\ket{\uparrow}+\sqrt{\frac{1}{1+\alpha}}\ket{\rightarrow}\nonumber\\
&=&\frac{1+\sqrt{\alpha}}{\sqrt{2+2\alpha}}\ket{\nearrow}+\frac{1-\sqrt{\alpha}}{\sqrt{2+2\alpha}}\ket{\searrow}.
\label{eq:psia}
\ea

This change of the wave function is an objective, physical event. Measuring the photon's $y$ polarization before and after the partial measurement will show that, at the statistical level, the $y$ polarization has been disrupted proportionately to the knowledge we gained about the $x$ polarization.

\section { Partial Measurement exerts a Partial Effect on Non-commuting Variables}

Let us now turn to the way partial measurement obeys the uncertainty principle. The effect of a partial measurement of $P_x$ on $P_y$ can be regarded as a rotation of the polarization plane: Drawing a vector with the $\ket{\rightarrow}$ component as the $x$ ordinate and the $\ket{\uparrow}$ component as the $y$ ordinate, will give $\theta=\tan^{-1}\Bigl(\frac{\braket{\uparrow}{\!\Psi_\alpha}}{\braket{\rightarrow}{\!\Psi_\alpha} }\Bigr)$ as the angle of the polarization plane. The initial state of $\ket{\nearrow}$ has two equal components of the $x$ polarization, namely, $\ket{\rightarrow}$ and $\ket{\uparrow}$, resulting in $\theta=45^{\circ}$. When a partial $P_x$ measurement is taken, the $\ket{\uparrow}$ component diminishes while the $\ket{\rightarrow}$ component increases, causing the polarization plane to rotate clockwise, until a complete measurement gives a pure $\ket{\rightarrow}$ state (Chart 1).

\begin{figure}[htbp]
\onecolumn\twocolumn[
\centering
\includegraphics{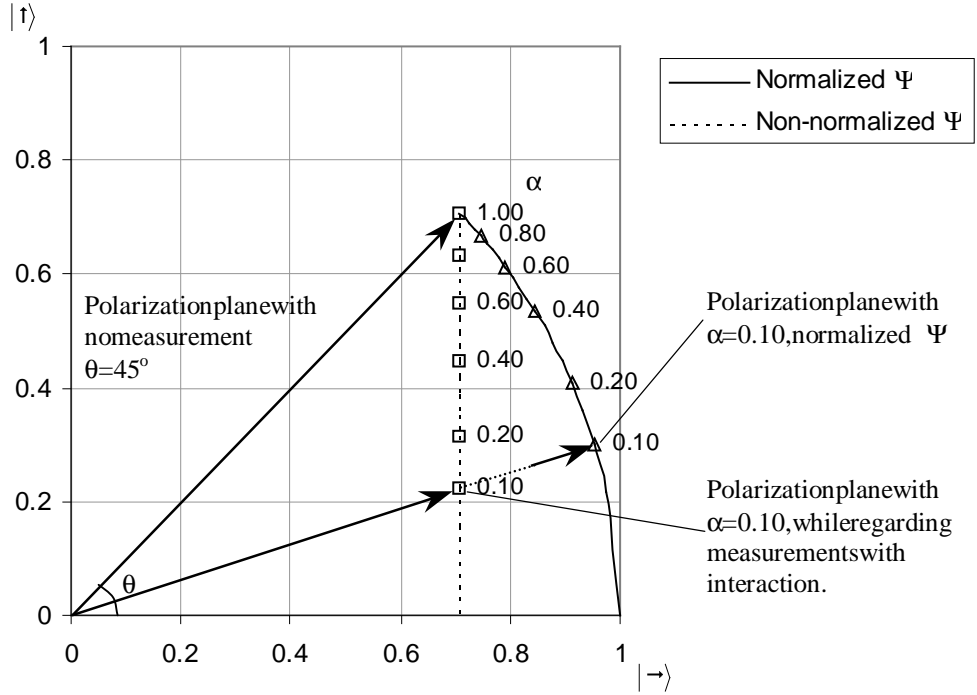}
\renewcommand{\thefigure}{1}
\renewcommand{\figurename}{Chart}
\onecolumn
\caption{ The polarization angle as a function of $\alpha$.}
\twocolumn
]
\end{figure}

Chart 1 also allows us to look at the vectors without normalizing $\Psi$ for $\Vert\!\ket{\!\Psi_\alpha}\Vert^2=1$: The formulation given in Eq.\ (\ref{eq:psia}) keeps the polarization vectors normalized since it counts only the cases where $\ket{\nearrow}$ or $\ket{\searrow}$ were measured -- excluding all the cases where a complete measurement took place and ended up with a click in one of the $\ket{\uparrow}$ detectors. An efficient way to look at the wave-function is to keep track of these $\ket{\uparrow}$ measurements. In this case, the probabilities for $\ket{\nearrow}$ and $\ket{\searrow}$ will not sum to 1, and the vector in Chart 1 will get shorter as $\alpha$ diminishes. Formally:
\ba
\label{eq:psita}
\ket{\!\Psi'_\alpha}&=&\sqrt{\frac{\alpha}{2}}\ket{\uparrow}+\sqrt{\frac{1}{2}}\ket{\rightarrow}\nonumber\\
&=&\frac{1+\sqrt{\alpha}}{2}\ket{\nearrow}+\frac{1-\sqrt{\alpha}}{2}\ket{\searrow}.
\label{eq:psiat}
\ea
This fact will turn out to be important later.

We can now calculate the correlation coefficient between the initial and the final $y$ polarizations as a function of the magnitude of the partial measurement:
\be
C_{y(\alpha)}=\big\Vert\braket{\nearrow}{\!\Psi_\alpha}\big\Vert^2=\left({1+\sqrt{\alpha} \over \sqrt{2+2\alpha}}\right)^2.
\ee
$C_{y(\alpha)}$ ranges from 1 (when a total correlation is kept) to 0.5 (when we have no correlation at all, since there is an equal probability to find $\ket{\nearrow}$ or $\ket{\searrow}$). For example, if measurements have been carried out on 50 out of the 100 $\ket{\uparrow}$ beams, yielding no click, then $\alpha=0.5$ and the agreement between the $y$ polarization before and after the $P_x$ measurement will be:

\[ C_{y(\alpha)}=\Bigl(\frac{1+\sqrt{\alpha}}{\sqrt{2+2\alpha}}\Bigr)^2 \simeq 97\%. \]
Indeed, since the $x$ and $y$ polarization directions are noncommuting variables, they satisfy the uncertainty relations mentioned in Eq.\ (\ref{eq:uncer}), and when applied to the above case of partial measurement, 
\be
\Delta P_x={2\sqrt{\alpha} \over 1+\alpha}, \; \Delta P_y={1-\alpha \over 1+\alpha},
\ee
which vary with $\alpha$ as in Chart 2.

\begin{figure}[htbp]
\onecolumn\twocolumn[
\centering
\includegraphics{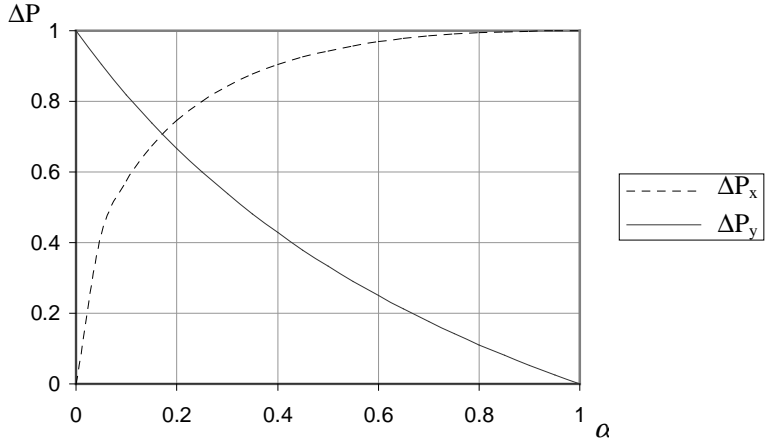}
\renewcommand{\thefigure}{2}
\renewcommand{\figurename}{Chart}
\onecolumn
\caption{ $\Delta P_x$ and $\Delta P_y$ as a function of $\alpha$.}
\twocolumn
]
\end{figure}

It is easily seen that as $\alpha$ increases (that is, fewer measurements on the $\ket{\uparrow}$ branch), $\Delta P_x$ also increases and $\Delta P_y$ decreases. That means that the more we measure on the $\ket{\uparrow}$ branch, the more precise knowledge we {\it gain} about $P_x$ (hence $\Delta P_x$ decreases), and the more knowledge we {\it lose} about $P_y$ (hence $\Delta P_y$ increases).

\section { Partial Measurement is Amenable to Complete Erasure}

An intriguing peculiarity of partial measurement is that, in contrast to the ordinary one, it can sometimes be totally reversed. To do this, there is no need to time-reverse the operation of any detector; one can merely repeat the partial measurement on the photon's opposite branch (Fig.\ 5). 

For example, let $n$ detectors be placed on $n$ paths of a photon's $\ket{\uparrow}$ branch. On average, in $\frac{200-n}{200}$ of the cases, these detectors will not click. This is a partial measurement, its outcome being given by Eq.\ (\ref{eq:measn}). Now let another battery of $n$ detectors be placed on $n$ paths of the same photon's $\ket{\rightarrow}$ branch. On average, in $\frac{200-2n}{200}$ of the cases, none of these detectors will produce a click either. This will completely undo the measurement and turn the wave-function back to the initial superposition:
\ba
&&\quad\frac{1}{\sqrt{2}}\left(\ket{\uparrow}+\ket{\rightarrow}\right)\longrightarrow\nonumber\\
&&\hbox{\it Measurement}\nonumber\\
&&\quad\;\; \longrightarrow\sqrt{100-n \over 200-n}\ket{\uparrow}+\sqrt{100 \over 200-n}\ket{\rightarrow}\nonumber\\
&&\quad\;\;\;=\sqrt{\alpha \over 1+\alpha}\ket{\uparrow}+\sqrt{1 \over 1+\alpha}\ket{\rightarrow}\longrightarrow\nonumber\\
&&\hbox{\it Erasure}\nonumber\\
&&\quad\;\; \longrightarrow\sqrt{100-n \over 200-2n}\ket{\uparrow}+\sqrt{100-n \over 200-2n}\ket{\rightarrow}\nonumber\\
&&\quad\;\;\;=\frac{1}{\sqrt{2}}\left(\ket{\uparrow}+\ket{\rightarrow}\right).
\ea
According to the notation we used for the partial measurement, the erasure process will take the form:
\be
\P{\rightarrow \alpha}\cdot \P{\uparrow \alpha}=\hat 1.
\ee
This reversal occurs when the two partial measurements on the opposing branches are of the same magnitude. In the more general case, of placing $n$ detectors on the $\ket{\uparrow}$ branch and $m$ detectors on the $\ket{\rightarrow}$ branch, the correlation (or rather the mismatch) between the initial and final $P_y$ measurement will be: 
\ba
\ket{\!\Psi_{\alpha\beta}}&\equiv&\P{\rightarrow\beta}\cdot\P{\uparrow\alpha}\ket{\!\Psi}\nonumber\\
&=&\sqrt{\alpha \over \beta+\alpha}\ket{\uparrow}+\sqrt{\beta \over \beta+\alpha}\ket{\rightarrow}\nonumber\\
&=&{\sqrt{\beta}+\sqrt{\alpha} \over \sqrt{2\beta+2\alpha}}\ket{\nearrow}+{\sqrt{\beta}-\sqrt{\alpha} \over \sqrt{2\beta+2\alpha}}\ket{\searrow},
\label{eq:psiab}
\ea
with correlation coefficient:
\be
C_{y(\alpha\beta)}=\left({\sqrt{\alpha}+\sqrt{\beta} \over \sqrt{2\alpha+2\beta}}\right)^2=\left({1+\sqrt{k} \over \sqrt{2+2k}}\right)^2,
\ee
where $\alpha$ and $\beta$ are the beam intensities of the $\ket{\rightarrow}$ and the $\ket{\uparrow}$ branches, respectively, and $k$ is the ratio between them:
\be
\alpha=\frac{100-n}{100}, \; \beta=\frac{100-m}{100}, \; k=\frac{\alpha}{\beta}.
\label{eq:abk}
\ee

Notice that $C_{y(\alpha\beta)}$ depends on the ratio $k$ alone. That means, for example, that a measurement of 50\% of the $\ket{\uparrow}$ branch ($k=\frac{1}{\sqrt{2}}$) will yield exactly the same $C_{y(\alpha\beta)}$ as a measurement of 90\% of the $\ket{\uparrow}$ branch and 80\% of the $\ket{\rightarrow}$ branch, or 99\% of the $\ket{\uparrow}$ branch and 98\% of the $\ket{\rightarrow}$ branch, etc. In other words, the exact number of paths on either branch does not matter, nor does their relative intensities, nor the number, nor the identity of the measured paths. The only thing that counts is the ratio of the {\it unmeasured} parts of the two branches!
\be
\P{\rightarrow\beta}\cdot \P{\uparrow\alpha}= \P{\uparrow\alpha/\beta}.
\ee
Consequently, the partial measurement operators in the $x$ direction are commutative:
\be
\P{\rightarrow\beta}\cdot \P{\uparrow\alpha}= \P{\uparrow\alpha}\cdot\P{\rightarrow\beta}=
\P{\uparrow\alpha/\beta} = \P{\rightarrow\beta/\alpha}.
\label{eq:commutative}
\ee

This ratio looks natural when discussing a single photon, as interference is known to depend on the intensities of the various beams diverging and re-converging from the initial wave function. Later, however, this ratio will prove to be of crucial significance as a proof for nonlocal influence between distant photons. 

\setcounter{figure}{4}
\begin{figure}[htbp]
\onecolumn\twocolumn[
\centering
\includegraphics{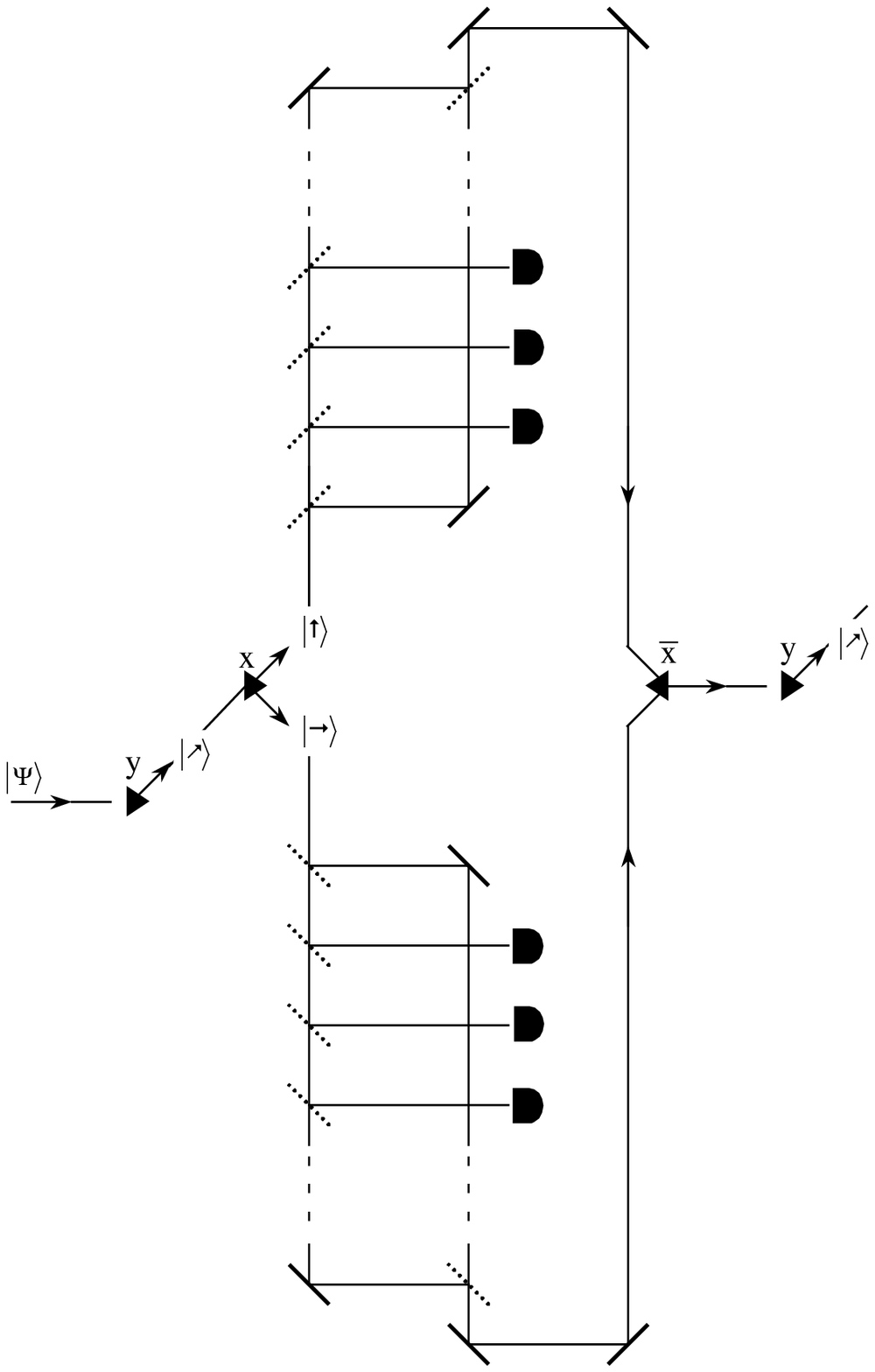}
\onecolumn
\caption{ When the same number of paths is interaction-freely measured on both the $\ket{\rightarrow}$ and $\ket{\uparrow}$ branches, their overall effects cancel each other and the photon returns to the original superposition.}
\twocolumn
]
\end{figure}

An important feature of the erasure process is that it complies with the ``no free lunch" principle. Any measurement on the $\ket{\rightarrow}$ path to restore the exact $\ket{\nearrow}$ state will also diminish an equal part of the $\ket{\uparrow}$ part, as seen in Fig.\ 6. To trace the cost of the erasure, we must keep track of the measured portion of the beam, hence a formulation similar to Eq.\ (\ref{eq:psiat}) is used with the inclusion of the erasure process:
\ba
\label{eq:psitab}
\ket{\!\Psi'_{\alpha\beta}}&=&\sqrt{\frac{\alpha}{2}}\ket{\uparrow}+\sqrt{\frac{\beta}{2}}\ket{\rightarrow}\nonumber\\
&=&\frac{\sqrt{\beta}+\sqrt{\alpha}}{2}\ket{\nearrow}+\frac{\sqrt{\beta}-\sqrt{\alpha}}{2}\ket{\searrow}.
\ea

\begin{figure}[t]
\centering
\includegraphics{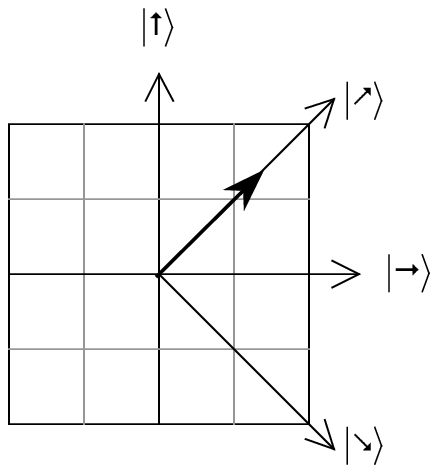}

{\small (a) Initial setup: $\ket{\!\Psi}=\ket{\nearrow}$.}

\vspace{10mm}
\includegraphics{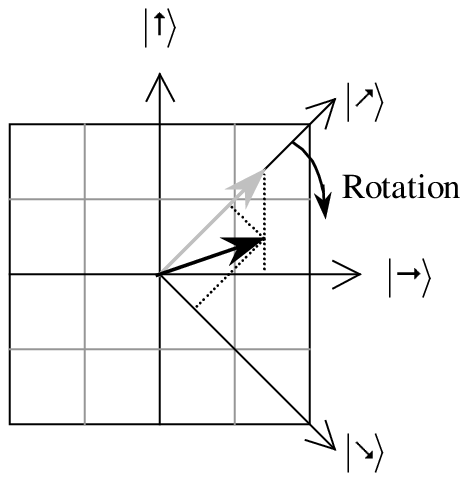}

{\small (b) A partial measurement on the $\ket{\uparrow}$ branch rotates the polarization plane: $\ket{\nearrow}$ diminishes while $\ket{\searrow}$ increases.}

\vspace{10mm}
\includegraphics{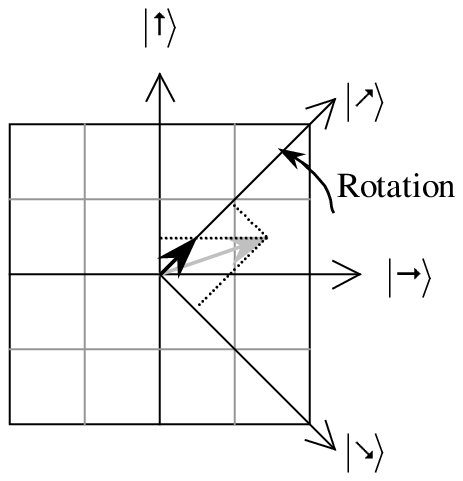}

{\small (c) Following a counter-measurement on the $\ket{\rightarrow}$ branch, the initial state is restored. The $\ket{\searrow}$ part is nullified but $\ket{\nearrow}$ diminishes too.}

\vspace{10mm}
\hspace{10pt}
\caption{ The cost of undoing a measurement.}
\end{figure}

\clearpage

Let us now summarize Sections II-IV: The splitting of the wave function into 200 paths enables carrying out a partial measurement, whose effect is manifested by the proportionate disruption of the interference effect. This setup also allows, in some cases, a complete erasure of the partial measurement and consequently a restoration of the interference pattern. 

\section{Nonlocal Effects of Partial Measurement and Erasure, Revealed by Interferometry}

Is it possible to show that such a partial measurement has nonlocal effects? Our proof involves an experiment with two particles in a singlet state. We show that, when both photons are subjected to interferometry, the partial measurement and erasure performed on each photon disrupt and restore, respectively, the interference effects of {\it both} photons.\footnote{For combining IFM and EPR experiments to prove the nonlocal nature of the former, see \cite{Ryff1,Ryff2}.}

\begin{figure}[htbp]
\onecolumn\twocolumn[
\centering
\includegraphics{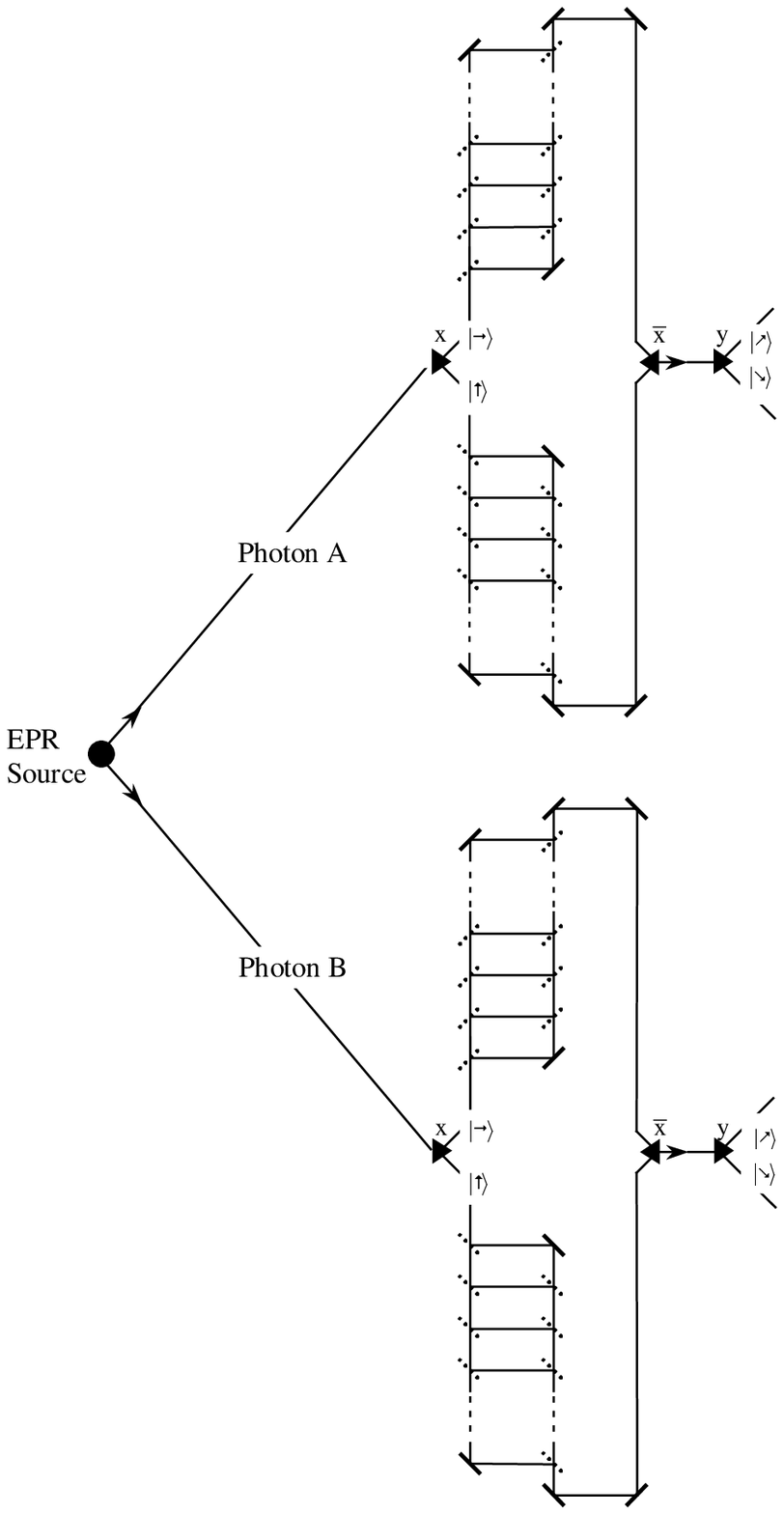}
\onecolumn
\caption{ An EPR-interferometry experiment. As long as no measurement of the two photons' $x$ polarizations is made, their final $y$ polarization will be 100\% correlated. When partial $P_x$  measurements are carried out, the changes in the correlation between their $y$ polarizations can demonstrate their nonlocal effects.}
\label{fig7}
\twocolumn
]
\end{figure}

Consider, then, a pair of spacelike-separated photons, A and B, in an entangled state, each entering an apparatus of the form described in Fig.\ 4. This is a hybrid EPR-PM experiment (Fig.\ \ref{fig7}). If both $P_x$ polarization are measured, they will be 100\% correlated but the $P_y$ polarizations will be unrelated. Conversely, if no detection is performed to find out which of the possible 200 paths any of the photons has taken, their $x$ polarization will remain unmeasured, hence their $P_y$ correlations will remain intact. 

Now let some detectors be placed on $n$ out of the 100 paths of the $\ket{\uparrow}$ branch of photon A. Photon B, in contrast, will not undergo any measurement of $P_x$, only of its $P_y$. 

In $\frac{200-n}{200}$ of the cases, all detectors measuring photon A will perform an interaction-free measurement, changing the wave function as in Eq.\ (\ref{eq:psia}). But here, due to the singlet state connecting the two photons, a unique state evolves. The partial measurement has partly disrupted the two photons' EPR entanglement: 
\ba
&&\ket{\!EPR}=\frac{1}{\sqrt2}\left(\ket{\uparrow}_1\ket{\uparrow}_2+\ket{\rightarrow}_1\ket{\rightarrow}_2\right)\nonumber\\
&&\quad \longrightarrow\sqrt{100-n \over 200-n}\ket{\uparrow}_1\ket{\uparrow}_2+\sqrt{100 \over 200-n}\ket{\rightarrow}_1\ket{\rightarrow}_2 \nonumber\\
&&\quad =\sqrt{\alpha \over 1+\alpha}\ket{\uparrow}_1\ket{\uparrow}_2+\sqrt{1 \over 1+\alpha}\ket{\rightarrow}_1\ket{\rightarrow}_2,
\label{eq:measepr}
\ea
where $\alpha$, again, is the intensity of the beam (as in Eq.\ (\ref{eq:alpha})). This change causes a decrease in the correlation between their $y$ polarizations:
\ba
\ket{\!\Psi_\alpha} &\equiv& \P{\uparrow_1\alpha}\ket{\!\Psi} \nonumber\\
&=&\sqrt{\alpha \over 1+\alpha}\ket{\uparrow}_1\ket{\uparrow}_2 +\sqrt{1 \over 1+\alpha}\ket{\rightarrow}_1\ket{\rightarrow}_2 \nonumber\\
&=&\frac{1+\sqrt{\alpha}}{2\sqrt{1+\alpha}}\left(\ket{\nearrow}_1\ket{\nearrow}_2+\ket{\searrow}_1\ket{\searrow}_2\right) \nonumber\\
&&\quad\quad + \frac{1-\sqrt{\alpha}}{2\sqrt{1+\alpha}}\left(\ket{\nearrow}_1\ket{\searrow}_2+\ket{\searrow}_1\ket{\nearrow}_2\right) \nonumber\\
&=&\frac{1+\sqrt{\alpha}}{\sqrt{2+2\alpha}}\ket{\!EPR}+\frac{1-\sqrt{\alpha}}{\sqrt{2+2\alpha}}\ket{\!\overline{EPR}}.
\label{eq:psiaepr}
\ea
where $\ket{\!\overline{EPR}}$ is the ``anti-EPR" state in which the photons are entangled, but in reverse polarization:
\be
\ket{\!\overline{EPR}}=\frac{1}{\sqrt{2}}\left(\ket{\nearrow}_1\ket{\searrow}_2+\ket{\searrow}_1\ket{\nearrow}_2\right).
\label{eq:psiepr}
\ee

\begin{figure}[htbp]
\onecolumn\twocolumn[
\centering
\includegraphics{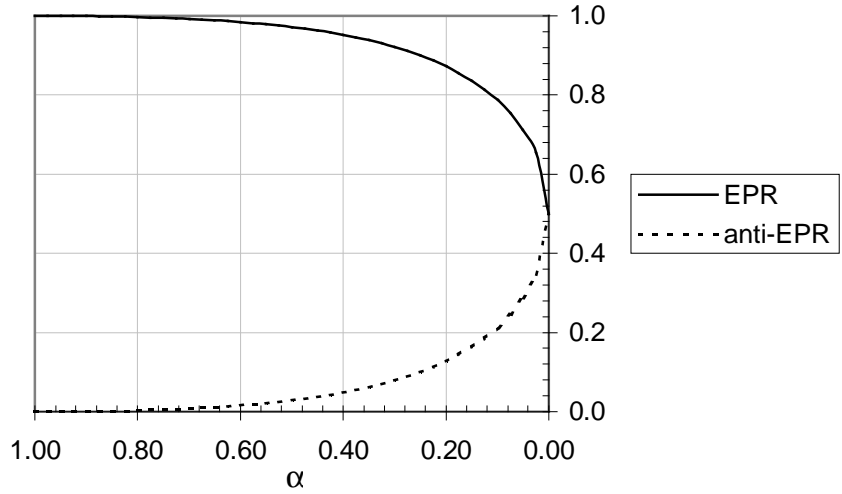}
\renewcommand{\thefigure}{3}
\renewcommand{\figurename}{Chart}
\onecolumn
\caption{ EPR and anti-EPR parts as a function of $\alpha$.}
\label{chart3}
\twocolumn
]
\end{figure}

As we can see, the partial measurement did not break the entanglement between the two photons. Instead, the pure EPR state became ``contaminated'' by a certain amount of the anti-EPR state. This means that measurements of the $y$ polarization of photons A and B have a probability of $\left({1+\sqrt{\alpha} \over \sqrt{2+2\alpha}}\right)^2$ to yield correlated results, and a probability of $\left({1-\sqrt{\alpha} \over \sqrt{2+2\alpha}}\right)^2$ to yield opposite results. The latter probability will grow as the magnitude of the partial measurements on $P_x$ grows (that is, as $\alpha$ diminishes) until the correlation drops to the random level of 50\%-50\% as $\alpha$ drops to 0 in case of a complete measurement (Chart \ref{chart3}).

As we will show below, the fact that the photons remain entangled even {\it after} partial measurement was performed enables us to increase or decrease the amount of the anti-EPR component in subsequent measurements.

Our next aim is to show that such a reversal can erase not only the outcome of a measurement performed on the same photon, but also that of the {\it other} photon. Consider the erasure of a partial measurement as described in Section \ref{se:partial}. If the erasure works, i.e., the measurement of the opposing branch also turns out to be interaction-free, the correlation between the two photons' $y$ polarizations will be restored: 
\ba
\ket{\!\Psi_{\alpha\beta}}&\equiv&\P{\rightarrow_1\beta}\cdot\P{\uparrow_1\alpha}\ket{\!\Psi} \nonumber\\
&=&\sqrt{\alpha \over \alpha+\beta}\ket{\uparrow}_1\ket{\uparrow}_2+\sqrt{\beta \over \alpha+\beta}\ket{\rightarrow}_1\ket{\rightarrow}_2 \nonumber\\
&=&\frac{\sqrt{\beta}+\sqrt{\alpha}}{2\sqrt{\alpha+\beta}}\left(\ket{\nearrow}_1\ket{\nearrow}_2+\ket{\searrow}_1\ket{\searrow}_2\right) \nonumber\\
&&\quad\quad +\frac{\sqrt{\beta}-\sqrt{\alpha}}{2\sqrt{\alpha+\beta}}\left(\ket{\nearrow}_1\ket{\searrow}_2+\ket{\searrow}_1\ket{\nearrow}_2\right) \nonumber\\
&=&\frac{\sqrt{\beta}+\sqrt{\alpha}}{\sqrt{2\alpha+2\beta}}\ket{\!EPR}+\frac{\sqrt{\beta}-\sqrt{\alpha}}{\sqrt{2\alpha+2\beta}}\ket{\!\overline{EPR}}.
\ea
Where $\beta$ is the intensity of the $\ket{\rightarrow}$ branch, as in Eq.\ (\ref{eq:abk}).

It is now clear that when $\alpha=\beta$, the $\ket{\!\overline{EPR}}$ part diminishes and the final state is the original $\ket{\!EPR}$ state, restoring the initial entanglement of photons A and B. 

Note that the restoration process is reminiscent of a procedure proposed by Deutsch {\it et al}. \cite{Deu} for ``entanglement purification" of EPR-like pairs. However, their procedure suffers from lack of information about the partially-entangled state. Consequently, it necessitates destroying every other transmitted particle, which serves as ``entanglement-control" particle, and then destroying also the accompanying particle, should the ``entanglement-control" particle indicate a non-entangled state. Our experiment, in contrast, does not require ``testing" the particles for entanglement. {\it Each and every} particle that is being ``caught" by the detector in the ``counter-measurement" is a non-entangled particle. Magically the QM formalism ensures that only the non-entangled particles will be caught by the detector! \footnote{This point will be elaborated on Section \ref{sec:free}.}

Therefore, when partial measurements and erasures are performed on two entangled photons, the local and nonlocal interpretations markedly differ:

\begin{description}
\item[Local Argument A:]\footnote{Proving nonlocal action is always difficult as adherents of locality often come up with very awkward yet not-impossible local mechanisms. Disproving such mechanisms is a tedious task,yet essential for a proof's completion. We therefore consider and disprove here all possible localist arguments.} Both the disruption of the correlation and its restoration are performed only by photon A's local interaction with the nearby detector, without affecting photon B whatsoever. 
\item[Nonlocal refutation:] The correlation between the two $y$ polarizations will be restored {\it even if we perform the partial $x$ measurement on photon A and the undoing of this measurement on photon B}. 
\end{description}

Here is the proof. Let $\gamma$ and $\delta$ denote the intensities of the $\ket{\rightarrow}$ and $\ket{\uparrow}$ branches of photon B, respectively. If both photons are subjected to partial measurements, the pair's state will be:
\ba
\label{eq:psiabcd}
\ket{\!\Psi_{\alpha\beta\gamma\delta}}&\equiv&\P{\rightarrow_2\delta}\cdot\P{\uparrow_2\gamma}\cdot\P{\rightarrow_1\beta}\cdot\P{\uparrow_1\alpha}\ket{\!\Psi}\\
&=&\sqrt{\alpha\gamma \over \alpha\gamma+\beta\delta}\ket{\uparrow}_1\ket{\uparrow}_2+\sqrt{\beta\delta \over \alpha\gamma+\beta\delta}\ket{\rightarrow}_1\ket{\rightarrow}_2\nonumber\\
&=&\frac{\sqrt{\beta\delta}+\sqrt{\alpha\gamma}}{\sqrt{2\alpha\gamma+2\beta\delta}}\ket{\!EPR}+\frac{\sqrt{\beta\delta}-\sqrt{\alpha\gamma}}{\sqrt{2\alpha\gamma+2\beta\delta}}\ket{\!\overline{EPR}}.\nonumber
\ea
Again, the entanglement between the photons is kept (though evolving into the unique state similar to Eq.\ (\ref{eq:psiepr})), and the correlation between the photons' $y$ polarizations would be: 
\ba
C_{y(\alpha\beta\gamma\delta)}&=&\left\Vert\braket{EPR}{\!\Psi_{\alpha\beta\gamma\delta}}\right\Vert  ^2\nonumber\\
&=&\left(\frac{\sqrt{\beta\delta}+\sqrt{\alpha\gamma}}{\sqrt{2\alpha\gamma+2\beta\delta}}\right)^2\nonumber\\
&=&\left(\frac{1+\sqrt{K}}{\sqrt{2 +2K}}\right)^2,
\label{eq:cyepr}
\ea
where, again, $\alpha$ and $\beta$ are, respectively, the beam intensities of the $\ket{\rightarrow}$ and the $\ket{\uparrow}$ branches of photon A, and $\gamma$ and $\delta$ are the $\ket{\rightarrow}$ and $\ket{\uparrow}$ branches of photon B. $K$ is the ratio between the intensities of the two photons: 
\be
K=\frac{\beta\delta}{\alpha\gamma}.
\ee
A few points are worth mentioning here:

\begin{enumerate}

\item The $\ket{\!EPR}$ component will always be greater than or equal to the $\ket{\!\overline{EPR}}$ part. Hence, in the above setup, one can never reach a situation when the two photons are manifestly anti-correlated.
\item In the extreme case when either $\alpha$, $\delta$, $\beta$, or $\gamma$ equals 0 (that is, a complete measurement was performed), the resultant state is an even blend of $\ket{\!\!EPR}$ and $\ket{\!\!\overline{EPR}}$ states, and $C_{y(\alpha\beta\gamma\delta)}=\frac{1}{2} $. That means that, following a complete measurement of $P_x$ on one photon, no information can be obtained about the other's $y$ polarization, as there are even probabilities to find the other photon in the same polarization ($\ket{\!EPR}$) or in the opposite one ($\ket{\!\overline{EPR}}$).

\item The opposite extreme case occurs when $\alpha\delta=\beta\gamma$, thereby $C_{y(\alpha\beta\gamma\delta)}=1$. Then, the two photons will restore the original $\ket{\!EPR}$ state of 100\% entanglement.
\item The process of ``erasing" the measurement bears a cost: Getting rid of the anti-EPR component will take the toll of diminishing the EPR part in equal amount. That means that many measurements will end up with a click. In order to measure this effect, we have to rewrite Eq.\ (\ref{eq:psiabcd}) without normalization for $\Vert\Psi\Vert^2=1$ after each measurement (as we did before in Eqs.\ (\ref{eq:psita}) and (\ref{eq:psitab})). When keeping $\ket{\!\Psi}$ relative to it's initial intensity (where $\alpha=\beta=\gamma=\delta=1$), we get: 
\ba
\ket{\!\Psi'_{\alpha\beta\gamma\delta}}&=&\frac{\sqrt{\beta\delta}+\sqrt{\alpha\gamma}}{2}\ket{\!EPR}\nonumber\\
&&\quad\quad +\frac{\sqrt{\beta\delta}-\sqrt{\alpha\gamma}}{2}\ket{\!\overline{EPR}}.
\ea
Now, for example, if a measurement of 50\% was taken on $\alpha$, the result will be: 
\be
\ket{\!\Psi'_1}=\frac{1+\sqrt{^1\!/\!_2}}{2}\ket{\!EPR}+\frac{1-\sqrt{^1\!/\!_2}}{2}\ket{\!\overline{EPR}},
\ee
whereas after a counter-measurement of 50\% on $\beta$ or $\delta$, the result will be: 
\be
\ket{\!\Psi'_2}=\frac{1}{2}\ket{\!EPR}.
\ee
Hence erasing the anti-EPR part took the toll of another reduction of $\frac{1}{\sqrt{2}}$ out of the EPR part. That is the reason why we cannot retrieve the EPR state after a {\it complete} measurement on one of the branches: The resultant state will be $\frac{1}{\sqrt{2}}{\nolinebreak\ket{\!EPR}}+\frac{1}{\sqrt{2}}\ket{\!\overline{EPR}}$. Trying to eliminate the anti-EPR component will also nullify the EPR component, leaving a fully measured photon.
\item Since $C_{y(\alpha\beta\gamma\delta)}$ depends on $K$ alone, the process is {\it inherently} non-local. $K$ is the ratio of the partial measurements on both particles and cannot be ``compensated'' on one particle without knowing the ratio of measurement on the other. This non-locality causes a Bell-like inequalities to break (see the refutation of Local Argument D in Section \ref{se:ineq} for a demonstration of such an example).

\end{enumerate}

Once, however, a pair of initially-entangled photons has survived the partial measurements of their $\ket{\rightarrow}$ and $\ket{\uparrow}$ branches with $K=1$ -- no matter whether the partial measurements were carried out on one photon or on both -- they restore their entanglement, hence the correlation in their $y$ polarizations. This offers a new extension of the EPR argument: 

\begin{quote}
{\it Just as quantum measurement imposes the measured polarization on the distant photon, so does quantum erasure obliterate the other photon's polarization.}
\end{quote}

Equation (\ref{eq:cyepr}) reveals another feature of the $P_y$ correlation between the photons: The combined effects of partial measurements and partial ``counter-measurements" do not comply with the ordinary subtraction rules. Consider, first, the case of a single particle. Its $y$ polarization is determined by the {\it relative} intensities of the $\ket{\rightarrow}$ and $\ket{\uparrow}$ branches, as in Eqs.\ (\ref{eq:psiab}) and (\ref{eq:abk}). To put it more pictorially, this is a natural consequence of the interference effect: If an interaction-free measurement has occurred, say, in 90\% of the $\ket{\uparrow}$ branch, and in 95\% of the $\ket{\rightarrow}$ branch, then the entire wave function resides in the remaining 10 and 5 percents, yielding a $K$ value of 2. Hence, the resulting $P_y$ correlation would be identical to that obtained by measuring just 50\% of the $\ket{\rightarrow}$ branch alone (giving again, $K=2$).

Now, when considering the interference effects of {\it two} such photons in the EPR setup, the nonlocality assumption yields another straightforward prediction that differs from the local assumption: 

\begin{description}

\item[Local Argument B:] The above deviation from ordinary subtraction rules stems from the interference effects occurring in each photon, regardless of what happens with the other photon. 

\item[Nonlocal refutation:] The singlet state obliges the above subtraction rules to equally
 hold even when the $\ket{\uparrow}$ branch is measured in photon A and the $\ket{\rightarrow}$ branch is measured in photon B. This effect is obliged by Eq.\ (\ref{eq:cyepr}), which shows that the $y$ polarization coefficient $C_{y(\alpha\beta\gamma\delta)}$ is a function of the measurement ratio $K$ alone.\footnote{Moreover, the restoration of the $P_y$ correlation cannot occur if we perform the undoing on {\it both} photons at the same time. By Eqs.\ (\ref{eq:measepr}) and (\ref{eq:psiaepr}), the surplus undoing constitutes a measurement in itself, which would disrupt again the correlation. For any partial measurement that has changed the initial superposition, we need only one ``counter measurement" of the same magnitude, {\it on either photon}, in order to restore it. This indicates that each measurement of one photon instantly affects the interference effects observed in the other photon.}

\end{description}

To summarize, in all cases in which interaction-free measurements are carried out on the opposing branches $\ket{\uparrow}$ and $\ket{\rightarrow}$, they mutually cancel out in the same way, {\it no matter whether they have been carried out on the same photon or on two entangled ones.} This indicates that each measurement effects the distant photon too.

\section { Introducing the Experimenter's Free Choice}
\label{sec:free}

Let us consider the next difference between quantum theory and the local prediction:

\begin{description}

\item[Local Argument C:] All the effects stem from a simple pre-established correlation between the photons, committing them to give the same results to the partial measurements.\footnote{Note that such a local account cannot be entirely classical. It cannot assume that the photon traverses only one out of the 200 paths, because in that case no interference will be observed. The local account must therefore go along the lines of the ``guide wave" interpretation. Nonetheless, as we show below, this account will not restore locality either.}

\end{description}

But such an argument enforces nonlocality in a new way: 

\begin{description}

\item[Nonlocal refutation:] In order for each single photon to be capable of responding to a certain number of detectors with silence, the photon must maintain a nonlocal connection between all the 200 parts of its wave function traversing distant paths. After all, the photon cannot know in advance on which paths the experimenter will choose to place the detectors!

\end{description}

This aspect of the experiment parallels the last-minute choice of the polarization direction in Aspect \& Grangier \cite{Asp} experiment or the GHZ \cite{Gre} experiment. A realization of the experiment would therefore require a random process for choosing the location and number of the $n$ out of the 100 paths to be measured. 

We have therefore proved that either the two photons maintain nonlocal correlation between them, or each photon maintains nonlocal correlation between its distant beams. The latter interpretation would join Hardy's \cite{Har3,Har4} and Albert {\it et al.}'s \cite{Alb} proofs for the nonlocality of a single photon. Either way, nonlocality is inescapable.

The setup in Fig.\ \ref{fig7} also points out the difficulty in applying counterfactuals to quantum mechanics: Suppose we measure 50 of the 100 $\ket{\uparrow}$ beams of photon A. A possible result of such an experiment might be that photon A finishes in the $\ket{\nearrow}$ path, while photon B goes on the $\ket{\searrow}$ path.

A possible description of such a situation will be: {\it There is a 3\% probability that the photons will be anti-correlated, and this is one of those cases.} But then, a counterfactual can be presented: What would have been the result had we placed detectors in front of 50 out of the 100 beams of the $\ket{\uparrow}$ branch of photon B too? The answer is intriguing: Since doing so will completely undo the measurement on photon A, photon B must either hit one of the 50 detectors or completely agree with the $y$ measurement of photon A. Since our photon did not agree with it, it must have been captured by one of the 50 detectors!

This imposes another odd counterfactual:
\begin{quote}
{\it When we place a battery of counter-measuring detectors on path B, they ``magically'' capture all the photons that {\em were about} to disagree with the $y$ measurement of photon A, have we not placed the counter-measuring detectors.}
\end{quote}
Such a teleological view is, of course, alien to physics. One must rather accept the objective reality of the wave-function. Only such a view can accept that the 50 detectors on photon B cancel exactly the 50 measurements done on photon A.

\section { An inequality for partial measurements}
\label{se:ineq}

Let us now give a general nonlocality proof for partial measurements. We shall consider a local hypothesis that tries to maintain locality despite the above predictions and show that it must violate an inequality theorem. 

\begin{figure}[htbp]
\onecolumn\twocolumn[
\centering
\includegraphics{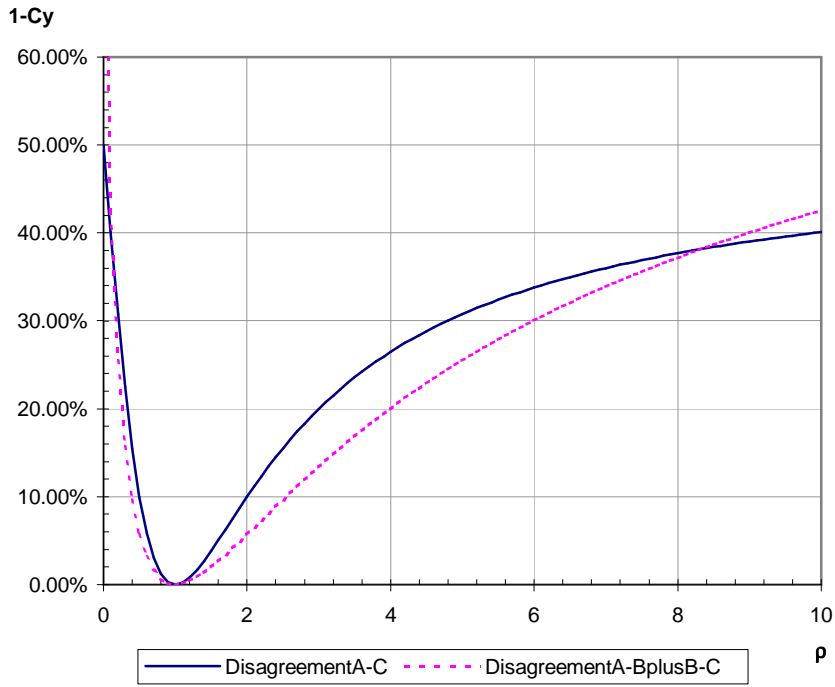}
\renewcommand{\thefigure}{4}
\renewcommand{\figurename}{Chart}
\onecolumn
\caption{ The disagreement between observers A and C is higher that the sum of disagreement between A and B plus B and C for a broad range of ratios, disproving locality by breaking a Bell-like inequality.}
\label{ch:refutation}
\twocolumn
]
\end{figure}

\begin{description}

\item[Local Argument D:] Each pair of photons uses a pre-established algorithm that assigns a definite $P_y$ value for each partial measurement: For any number of paths that were interaction-freely measured, the photons would yield some preestablished $y$ polarization. The resulting list of $P_y$ values is infinite, matching every possible degree of partial measurement. 

\item[Nonlocal refutation:] The alleged algorithm must satisfy two restrictions: (1) In every pair, both photons must obey the same algorithm (though in reverse polarities): if photon A undergoes a measurement of 30\% on its $\ket{\uparrow}$ branch, and photon B undergoes the same measurement of its $\ket{\rightarrow}$ branch, their $P_y$ measurement {\it must} agree on each single experiment ($K=1$, hence $C_{y(\alpha\beta\gamma\delta)}=1$). This, indeed,  explains the apparent ``erasure," where opposite partial measurements on the two photons restore the initial correlations. (2) However, the algorithm must assign the particular $y$ polarization to the {\it ratio} of the intensities of the photon's $\ket{\uparrow}$ and $\ket{\rightarrow}$ branches (that is $\frac{\alpha}{\beta}$ for photon A and $\frac{\gamma}{\delta}$ for photon B). The correlation ratio $C_{y(\alpha\beta\gamma\delta)}$ makes this fact evident: A measurement of 50\% on the $\ket{\uparrow}$ branch of photon A yields $\frac{\alpha}{\beta}=0.5$, but many other measurements on photon B will result $\frac{\gamma}{\delta}=0.5$ too (0/50, 60/80, 80/90, etc.), equating $C_{y(\alpha\beta\gamma\delta)}$ to 1 and restoring the photons' $P_y$ correlation. 

Now, restrictions (1) and (2) refute the nonlocal argument by breaking a Bell-like inequality in the following way: Consider an experiment where photon A has $\frac{\alpha}{\beta}=1.0$ and photon B has $\frac{\gamma}{\delta}=0.5$. Here, $K=0.5$, hence $C_{y(\alpha\beta\gamma\delta)}=0.97$, implying that  the $P_y$ correlation between the photons is disrupted in 3\% of the cases. If one believes in a pre-existing algorithm directing each photon, the following counterfactual must be true: Should B now had $\frac{\alpha}{\beta}=0.5$ and it repeated the same measurement, but with a different opponent, C, with square the measurement ratio: $\frac{\gamma}{\delta}=0.25$, the results {\it must} have been the same! Since B must give a $P_y$ measurement according to it's $\frac{\alpha}{\beta}=0.5$ ratio, it must measure exactly the same result that it gave for that ratio in the actual experiment (in accordance with restriction (1) -- A and B must obey the same algorithm -- and restriction (2) -- the algorithm depends on $\frac{\alpha}{\beta}$ alone). Since $K$ remains with the same value ($1/0.5=0.5/0.25=2$), $C_{y(\alpha\beta\gamma\delta)}$ remains 0.97, so B and C must give non-correlated results in 3\% of the cases too. This imposes another counterfactual: If A measured $\frac{\alpha}{\beta}=1.0$ against C with $\frac{\gamma}{\delta}=0.25$, they could give, at most, different results in $3\%+3\%=6\%$ of the cases. However, when we compute $C_{y(\alpha\beta\gamma\delta)}$ for this case ($K=4$), the result is 0.90. Which means that they {\it must} give non-correlated results in 10\% of the cases! Since $10\%>3\%+3\%$, that condition cannot be met, and we conclude that the local argument is false. Q.E.D. 

\end{description}

This proof will be generalized below for a broad range of ratios. That is, for a certain measurement ratio $\rho$, the disagreement between A and C is greater than the sum of disagreement between A and B plus B and C (see Chart \ref{ch:refutation}):

If the difference between measurement ratios of particles A and B is $K=\rho$, the disagreement in $P_y$ measurements will be:
\be
\Delta_{A-B}=1-C_{y(\alpha\beta\gamma\delta)}=1-\left(\frac{1+\sqrt{\rho}}{\sqrt{2 +2\rho}}\right)^2.
\ee
The difference between measurement ratios of B and C will be $\rho$ too, yielding the same value for $\Delta_{B-C}$ (since it depends on the ratio $\rho$ alone), hence the maximal sum of the disagreement between A and C will be twice the above amount:
\be
\Delta_{A-B}+\Delta_{B-C}=2-2\left(\frac{1+\sqrt{\rho}}{\sqrt{2 +2\rho}}\right)^2.
\ee
However, the measured disagreement between A and C will be according to $K=\rho^2$:
\be
\Delta_{A-C}=1-\left(\frac{1+\rho}{\sqrt{2 +2\rho^2}}\right)^2.
\ee

In Chart \ref{ch:refutation} we show the graphs of these functions, which shows that in the region $1<\rho<8.3$ the disagreement $\Delta_{A-C}$ is higher than the sum $\Delta_{A-B}+\Delta_{B-C}$, thereby disproving any possible local explanation. 

It should also be pointed out that Local Argument D is especially ludicrous when we consider its {\it post hoc} explanations for the unique quantitative features of joint interferometry. Why should the ratio 50\%-0\% of measurement and erasure give the same result as 75\%-50\%, 90\%-80\% and so on? A local model can ``explain" these phenomena only by adding arbitrary assumptions without any rationale other than the need to account for such unexpected results. In the nonlocal account, in contrast, these peculiarities are straightforwardly derived from {\it i)} the very nature of interference, and {\it ii)} the assumption that the two interferometries affect one another due to the quantum entanglement of the two photons.

Finally let the proof be extended to include all pairs:

\begin{description}

\item[Local Argument E:] Perhaps only those photons that give rise to partial measurements maintain nonlocal correlation, while the others, which react to the measurement with a click in the detectors (complete measurement), have pre-fixed correlation and do not affect one other nonlocally. These photons, in other words, have ``agreed" in advance to respond to the detectors with clicks, and therefore need not show correlation in their $y$ polarization.

\end{description}

The refutation of this hypothesis (see also Ref.\ \cite{Eli5}) is just like the proof we used in Section \ref{sec:free}: 

\begin{description}

\item[Nonlocal refutation:] In order for some photons to be capable of responding to a certain number of detectors with a click, each such a photon must maintain a nonlocal connection between the 200 distant parts of its wave function. For the photon cannot know in advance in which of the paths are detectors going to be placed. Therefore, {\it once the partial measurements confirm the nonlocal prediction, the complete measurements equally indicate nonlocal effects.}

\end{description}

\section { Multiple Partial Measurement}

Since partial measurement does not take the cost of disentanglement, it can be followed by many consecutive partial measurements, all of which affect the distant entangled particle. This is in marked contrast with the ordinary EPR experiment, which allows the nonlocal transfer of only one variable. 

Consider the following case: Of two entangled particles, one undergoes partial measurement of its $x$ polarization, then of its $y$ polarization, and then of its $z$ polarization\footnote{We will use the $z$ `direction' to denote a circular polarization. The $z$ direction is taken from the spin measurement of spin-$^1\!/_2$ particles which obeys the same relations as the $x$, $y$, and circular polarization measurement. The appropriate operators will be $\P{\odot\alpha}$ and $\P{\otimes\alpha}$.}. Suppose also that all partial measurements were very close to 100\% (say, 90\% each). Of course, such a multitude of measurements increases the probability for a click which would ruin the experiment, but if all partial measurements succeeded, the state of the two entangled particles is:
\be
\P{\odot_1 90\%}\cdot \P{\nearrow_1 90\%}\cdot \P{\uparrow_1 90\%}\cdot\ket{\!EPR}.
\ee

The resulting pair of particles is poorly correlated, but an inverted set of counter measurements, performed on the particle that was measured, can restore the initial correlation. Since measurements in orthogonal directions cannot commute, the counter measurements must be performed in a reverse order:
\ba
&&\P{\rightarrow_1 90\%}\cdot\P{\searrow_1 90\%}\nonumber\\
&&\quad \cdot(\P{\otimes_1 90\%}\cdot\P{\odot_1 90\%})\nonumber\\
&&\quad\quad \cdot\P{\nearrow_1 90\%}\cdot\P{\uparrow_1 90\%}\cdot\ket{\!EPR}\nonumber\\
&& = \P{\rightarrow_1 90\%}\cdot(\P{\searrow_1 90\%}\cdot\P{\nearrow_1 90\%})\cdot\P{\uparrow_1 90\%}\cdot\ket{\!EPR}\nonumber\\
&& = (\P{\rightarrow_1 90\%}\cdot\P{\uparrow_1 90\%})\cdot\ket{\!EPR}\nonumber\\
&& = \ket{\!EPR}.
\ea
In general, any set of partial measurements gives a complex linear combination of the kind:
\ba
\ket{\!\Psi}&=&(a+ib)\ket{\uparrow}_1\!\ket{\uparrow}_2+(c+id)\ket{\rightarrow}_1\!\ket{\rightarrow}_2\nonumber\\
&&\quad+(e+if)\ket{\uparrow}_1\!\ket{\rightarrow}_2+(g+ih)\ket{\rightarrow}_1\!\ket{\uparrow}_2.
\ea
A pair of particles, therefore, can carry up to eight real parameters. Such a combination of states, once successfully created on one particle, shows up in the distant particle too, a case that is impossible in the ordinary EPR experiment. In terms of quantum information, this state far exceeds the present, single-bit nonlocal correlation.

Let us note another interesting peculiarity of multiple partial measurement: If one particle has undergone a series of partial measurements, then the erasure of these measurements can be attempted only in the reverse order (last first, first last). If, however, the erasure is attempted on the other particle of the entangled pair -- even after a time-like interval -- the order of the erasures must be of that of the measurements (first first, last last). This fact seems to lend support to Cramer's \cite{Cra} ``transactional interpretation," where the measurement of one particle affects the other particle by traversing a ``Feynman zigzag" through time. 

\section { The New Quantum Erasure and its Significance}

``Quantum erasure" denotes an operation that constitutes the time reversal of the measurement process, undoing the the measurement's outcome and turning the wave function back to its initial pure state. It has become the focus of intensive study during the last few years because of its far-reaching theoretical and technological bearings, such as quantum computation and reversibility. 

Most notable of these works is that of Scully {\it et al.}, \cite{Scu} who demonstrated erasure of a quantum measurement in a double-slit experiment with atoms. In this experiment, the two parts of the wave function, prior to being reunited, pass through a small cavity where they undergo a measurement that can tell which path the atom has traversed. Immediately after the measurement, while the beams are still inside the cavities, the ``which path" information is totally erased. Scully {\it et al.} showed that after leaving the cavity, the two beams gave rise to normal interference, just as if no measurement took place.

This experiment, however, has two shortcomings that obscure the uniqueness of quantum erasure. First, the experiment makes it impossible to know the actual result of the measurement that has been later erased. This impossibility is imposed by the very definition of the experiment: If an experimenter observes the result of a measurement, this observation itself becomes part of the measurement, hence erasure requires completely erasing that observer's brain processes as well. Therefore, one can only {\it infer} that detection and its erasure took place on one of the two beams, but never know which beam gave rise to the initial click. The same holds for the proposals of Greenberger \& YaSin, \cite{Gre2} Becker, \cite{Bec} and others, reviewed and refined by Kwiat {\it et al.} \cite{Kwi2}.

Our proposal overcomes this limitation. When the measurement is partial, it is as observable as any other measurement, and similarly its erasure. The reason for this has been pointed out earlier: Unlike the prevailing techniques, ours does not require time-reversing any measuring instrument, thereby avoiding the enormous technical difficulties involved with proper atomic control and thermodynamic isolation. Rather, the only thermodynamic price we have to pay is that, the closer the measurement gets to a complete measurement, the more likely it is to end up in a click, whereby erasure will be no more possible. Similarly, erasure itself might end up in a click. We thus comply with the Second Law of Thermodynamics at the statistical level. Still, in a large enough series of experiments, we can have many cases in which we can directly observe a measurement that is close enough to complete measurement, and then observe its complete erasure. 

But the most intriguing consequence of quantum erasure is obscured by the second shortcoming of Scully {\it et al.}'s experiment: They carried out the measurement and its erasure on {\it both} halves of the wave function. This, we suggest, is not only unnecessary but suppresses the nonlocal aspect of the process. Elitzur \cite{Eli2} has proposed to repeat Scully's experiment dropping one of the measurement and its erasure on one side. The argument was that if interference shows up in this case too, it would prove that undoing one half of the wave function instantly ``unmeasures" the other half too. However, this proposal suffers from the same shortcoming as the above erasure experiments: The measurement and its erasure can only be inferred and never directly observed. Brun and Barnett, \cite{Bru} on the other hand, proposed to carry out the measurement on one arm and the cancellation on the other, arguing that both operations should affect both arms. 

Unfortunately, all these experiments study single-particle interference, where the two parts of the wave function are eventually reunited. This does not allow nonlocality to be tested. A local theory could argue that both the measurement and its erasure affect only the measured half. 

Partial measurement, again, overcomes all these shortcomings. It allows a direct observation of both the measurement and its erasure. Furthermore, by performing the measurement on one photon and the erasure on the other, and by {\it not} uniting the two distant photons, we are in a position to affirm: {\it Not only quantum measurement, but also its erasure, affects the other particle in the EPR experiment}. Notice also that, unlike the micromaser cavities and laser beams in Scully {\it et al.}'s experiment, our method carries out the erasure by much simpler optical means.

\section {Conceptual Implications}

In a discussion that has become a classic, Feynman \cite{Fey} defined the double-slit experiment as ``the core of the mystery of quantum mechanics." Albert \cite{Alb2} has generalized this insight in his lucid exposition of quantum mechanics, where he showed that any quantum-mechanical variable is a superposition of another, noncommuting one. He went on to show how any variable could emerge from the interference effect of its conjugate. 

This is what the present work does with the $x$ and $y$ polarizations, illustrating the uncertainty principle by the familiar phenomenon of interference. It is this highly visualizable phenomenon upon which the present proof for nonlocality is based, not requiring Bell's theorem. In so doing, we have also extended the EPR argument to two conclusions: {\it i)} Nonlocal effects are caused not only by a complete measurement but even by its minutest stages. {\it ii)} Nonlocal effects are caused not only by a measurement but also by the time-reversed process. Therefore, nonlocality needs not cease once the two particles in the EPR experiment are measured; they can remain entangled in spite of many successive measurements, maintaining the strange ``dialogue" between them for a long time. 

Our proof also extends Bell's proof in that, whereas Bell's inequality is based on the polarization measurements along varying angles, we present a new inequality that holds within the same angle of polarization for both particles. 

Partial measurement gives a new twist to the question whether God plays dice. Our experiment shows that not only does God cast a dice every time a polarization measurement is performed, but that, when the dice takes some time to fall, God preserves the right to change His mind as many times as He pleases. He can, for example, give 99\% for a particle to have a $\ket{\uparrow}$ polarization, and then, upon the $100^{th}$ partial measurement, change His mind and make the polarization $\ket{\rightarrow}$. Give Him further opportunities by dividing the $100^{th}$ measurement into further 100 partial measurements, and He might change his mind time and again. \footnote{This bearing of partial measurement on the issue of determinism {\it vs.} indeterminism makes it also relevant to the origins of time-asymmetry, see \cite{Eli3,Eli4,Eli41}.}

To summarize, our proof highlights a fundamental peculiarity of quantum mechanics. Suppose that a classical object resides in one out of 200 closed boxes. Opening some boxes and not finding the object there only alters the observer's {\it knowledge} (or, better, ignorance) about the object's location. In quantum mechanics, in contrast, every such non-detection brings about a real change in the particle's state, empirically proved in our EPR-PM setting with utmost quantitative precision. It is this lack of differentiation between ontology and epistemology -- any change in the observer's {\it knowledge} corresponding to precisely the same change in the {\it state} of the thing observed -- that makes quantum mechanics so unique among all natural sciences.

\section*{Acknowledgments}

An early version of this work was presented at the Gordon Conference on Modern Developments in Thermodynamics, Ventura, USA, February 1997. Thanks are due to the participants for their helpful comments. It is also a pleasure to thank Yakir Aharonov, Daniel Rohrlich and Lev Vaidman for illuminating discussions.


\end{document}